# Deflating the Spacetime–Matter Dichotomy

ANTONIO FERREIRO
ALEX FLEUREN
NIELS C. M. MARTENS
*Author affiliations can be found in the back matter of this article

## ABSTRACT

In this paper, we analyze scalar–tensor theories—specific instances of which include mainstream inflation and dark energy models—in light of the spacetime–matter (STM) dichotomy. We argue that it is difficult to categorize the scalar fields as either a pure aspect of the spacetime structure or a pure form of matter, by focusing on the Jordan vs. Einstein frames of these theories. We present and evaluate various interpretational options available, concluding that the STM dichotomy becomes untenable in this context. At the same time, the ontological and conceptual category of spacetime can be decoupled from that of gravity, with the latter remaining viable in the context of scalar–tensor theories.

CORRESPONDING AUTHOR:

**Antonio Ferreiro**

Freudenthal Institute, Utrecht University, NL; Descartes Centre for the History and Philosophy of the Sciences and the Humanities, Utrecht University, NL

a.e.ferreirodeaguiar@uu.nl

# 1 INTRODUCTION

Space(time) is a container, an arena. Matter is that which is contained—the actors dancing on the stage, the gladiators fighting within the Colosseum. When physicists postulate a new entity, say a new field, there is no way around it: it must be either an aspect of (i.e., a modification of) spacetime, say of its gravitational structure—a stage, part of the decor, the purview of the prop department—or a matter field—a bachatero, a ballerina. Or so the standard story goes.

Moving to field theories such as general relativity (GR) reveals the first crack in this overly simplistic dichotomy. The metric field, typically presented as the representative of spacetime, is dynamical like the familiar matter fields. It reacts when acted upon, like the other matter fields (Rovelli 1997). What remains controversial is whether it exhibits the final[1] hallmark of matter that would make it an equal dance partner of fields such as the electromagnetic field: carrying energy. The phenomenological tradition of Feynman, Bondi, Rovelli, and others (Feynman 1995; Rovelli 1997) claims that the metric does carry (gravitational) energy. Technical approaches focus on the pseudo-tensorial nature of the putative gravitational energy (Duerr 2019a,b; Hoefer 1996), and take this to preclude the metric from carrying genuine energy.[2] Only some highly symmetric models of GR, such as Schwarzschild solutions with their foliation-independent, constant ADM energy, may avoid the abovementioned worries (Read 2018); see Vergouwen and Martens (ms) for the nuanced considerations that come into play when attempting to categorize such solutions as being spacetime and/or matter solutions. The exact status, then, of the spacetime–matter (STM) dichotomy in the context of GR is controversial.

Further case studies, from astronomy and cosmology, provide fertile ground for analyzing the conceptual difference and mutual exclusiveness, if any, between spacetime and matter. Earlier work considered (a specific model of) dark matter (Martens and Lehmkuhl 2020a), which was argued to provide a stronger case for a breakdown of the STM distinction than GR, opening up various interpretational options (Martens and Lehmkuhl 2020b). Here we focus on scalar–tensor theories, of which various mainstream inflation and dark energy models are specific instances, e.g., Brans–Dicke theory, dilaton gravity and $F(R)$ gravity (a generalization of Starobinsky inflation).

Scalar–tensor theories merit careful analysis for at least two physical reasons. First, recent empirical developments have renewed their relevance in both cosmology and high-energy theory: they have been shown to provide viable frameworks for explaining late-time cosmic acceleration in the context of dark energy models (Wolf et al. 2025) and to accommodate inflationary dynamics compatible with current observational constraints (Kallosh and Linde 2025) Second, from a theoretical standpoint, scalar–tensor frameworks naturally arise in the consistent quantization of scalar fields in curved spacetime, even in cases where no non-minimal coupling is postulated at the outset—a fact emphasized in the standard treatment of quantum field theory in curved spacetime (Birrell and Davies 1982; Parker and Toms 2009).

The philosophical upshots of our investigation of whether the scalar field (e.g., the inflaton field or dark energy field) is to be categorized as spacetime, or matter, or both, or neither, are threefold. By applying well-known changes of variables—a conformal transformation of the metric combined with a redefinition of the scalar field—one moves between different representations of a scalar–tensor theory, the well-known Jordan frame and Einstein frame.

---

1 One stronger criterion would be that of having rest mass, but this is arguably too strong as it would rule out photons as being matter (Martens and Lehmkuhl 2020a).

2 Read (2018) attempts to bridge these two traditions by invoking functionalism.

The philosophical significance of such frame transformations in Brans–Dicke theory was already noted by Weinstein (1996), who argued that conformally related formulations can leave it unclear which metric should be identified with spacetime structure: the metric appearing in the gravitational field equations or the metric with respect to which matter propagates. On the standard interpretation of such transformations as indeed relating merely different descriptions of the same underlying physical state of affairs, our first point will then be that this causes problems for a black-and-white STM dichotomy since the categorization of the scalar field is not invariant across these representations. Second, we will see that even within each of the two representations the black-and-white dichotomy fails to apply, albeit for different reasons in each representation. A third reason why these scalar–tensor theories are conceptually and philosophically interesting is that, against this negative backdrop of a breakdown of an old dichotomy, the conceptual category of gravity, if decoupled from that of spacetime, comes to the fore as a renewed category that does apply, since the scalar field is consistently classified as a gravitational field across representations.

The structure of the paper is as follows. In Section 2, we introduce various criteria for being matter and being spacetime, respectively. In Section 3, we formally introduce scalar–tensor theories, and distinguish the Jordan and Einstein representations of such theories. The scalar field is then analyzed in each frame, i.e., the criteria from the previous section are used to determine whether the scalar field in each representation is spacetime, or matter, or both, or neither. Spoiler alert: the scalar field is neither spacetime nor matter in the Jordan representation; it is spacetime but not matter in one Einstein representation and matter but not spacetime in another version of the Einstein representation. We then present and evaluate the various philosophical strategies available to interpret this tension (Section 4)—does the gladiator win their freedom to leave the Colosseum? Yes! We conclude that, at least, the spacetime and matter categorization becomes a matter of convention, but the most honest exit route is to give up on a mutually exclusive dichotomy between spacetime and matter altogether. On the positive side, it turns out to be possible to decouple the spacetime category from the gravity category—the latter one does remain viable in the context of scalar–tensor theories.

## 2 STM DICHOTOMY IN RELATIVISTIC (METRIC) FIELD THEORIES

In this section, we examine relativistic metric field theories with the aim of (determining the tenability of) developing a precise framework for separating spacetime from matter. We begin by reviewing standard examples such as electromagnetism, perfect fluids, and GR, and by showing that these theories often admit multiple mathematically equivalent representations. While such representational freedom is unproblematic in simple cases, it poses serious challenges when we turn to scalar–tensor theories, where the STM distinction becomes blurred.

To address this, we formulate explicit criteria in an attempt to classify the fields of any given representation. Matter fields are required to have their own dynamical equations, to couple directly or indirectly with other matter fields, and to possess a conserved stress-energy tensor. Spacetime, by contrast, is identified with the manifold, the metric, and any fields that enter into the construction of the metric.[3] The traditional hope is that this yields a classification scheme that unambiguously decides for each element of the theory whether it is either spacetime or matter (i.e., an exclusive "or").

---

3 For a classic discussion of whether the manifold without any metric field(s) would suffice, see (Earman and Norton 1987; Hoefer 1996; Maudlin 1988).

We demonstrate the applicability of this framework in concrete examples, such as fluids interacting with electromagnetic or gravitational fields. In each case, the dichotomy holds: matter carries and exchanges energy–momentum, while spacetime supplies the geometric background. We conclude that the proposed classification provides a reliable method for preserving the STM dichotomy in relativistic field theories, and sets the stage for assessing whether this dichotomy remains viable in the context of scalar–tensor theories.

## 2.1 RELATIVISTIC (METRIC) FIELD THEORIES

For our purpose, a relativistic field theory can be understood as a family of kinematically possible models (KPM) of the form $\langle M, \psi_1, \ldots, \psi_n \rangle$, which are then restricted by dynamical equations to the dynamically possible models (DPM):

$$\langle M, \psi_1, \ldots, \psi_n \rangle \quad (+) \quad \text{Dynamical equations,} \tag{RFT}$$

where $M$ is a differentiable manifold, the $\psi_i$ are a finite number of fields,[4] and the dynamics are governed by a set of partial differential equations involving these fields.

In the case of metric field theories, we further require the existence of a Lorentzian metric tensor field $g_{ab}$, which may or may not be constructed from one or more of the fields $\psi_i$. Some standard examples of relativistic metric field theories include:

**(1.)** Maxwell's electromagnetism:

$$\langle M, \eta_{ab}, A_a, j^b \rangle \quad (+) \quad \nabla_a F^{ab} = j^b, \quad \nabla_{[a} F_{bc]} = 0, \tag{MAXW}$$

where $F^{ab} = \eta^{ac}\eta^{bd}F_{cd}$, $F_{ab} = \nabla_a A_b - \nabla_b A_a$ and $j^b$ is an external current, required only to satisfy the conservation equation $\nabla_b j^b = 0$. The covariant derivative $\nabla_a$ is defined with respect to the flat metric $g_{ab} = \eta_{ab}$.

**(2.)** Perfect fluid:

$$\langle M, \eta_{ab}, \rho, P, u^b \rangle \quad (+) \quad \nabla_a(\rho u^a) + P\,\nabla_a u^a = 0, \quad (\rho + P)u^b \nabla_b u^a + h^{ab}\nabla_b P = 0, \tag{PF}$$

where $\rho$, $P$, and $u^b$ represent the fluid's density, pressure, and four-velocity, respectively, and $h^{ab} = \eta^{ab} + u^a u^b$. Again, $\nabla_a$ is associated with the flat metric $\eta_{ab}$.

**(3.)** General relativity:[5]

$$\langle M, g_{ab}, T^{ab} \rangle \quad (+) \quad G^{ab} := R^{ab} - \tfrac{1}{2} R g^{ab} = T^{ab}, \tag{EFE}$$

where $g_{ab}$ is a Lorentzian metric with associated torsion-free and metric-compatible covariant derivative $\nabla_a$, $R^{ab}$ is the Ricci tensor of $g_{ab}$, $R$ the corresponding Ricci scalar, and $T^{ab}$ is the symmetric stress-energy tensor satisfying $\nabla_a T^{ab} = 0$.

In all these cases, and in the rest of the paper, we assume that the derivative operator appearing in the dynamical equations is uniquely determined by $g_{ab}$, via the compatibility condition $\nabla g = 0$.[6] The particular choice of fields $\psi_1, \ldots, \psi_n$ used to express a theory is known as a *representation* (Sotiriou, Faraoni and Liberati 2008; Thorne, Lee and Lightman 1973). For instance, we may reformulate (MAXW) as:

$$\langle M, \tilde{g}_{ab}, A_a, \tilde{j}^b, \phi \rangle \quad (+) \quad \tilde{\nabla}_a \tilde{F}^{ab} = \tilde{j}^b, \quad \tilde{\nabla}_{[a} F_{bc]} = 0, \tag{MAXW'}$$

where now $\tilde{F}^{ab} = \tilde{g}^{ac}\tilde{g}^{bd}F_{cd}$, $\tilde{g}_{ab} = \phi^2 \eta_{ab}$ and $\tilde{j}^b = \phi^{-4} j^b$ (Côté, Faraoni and Giusti 2019). Equations (MAXW) and (MAXW') are different representations of the same theory. A similar transformation can be applied to (PF). An alternative representation of the perfect fluid theory is:

$$\langle M, \tilde{g}_{ab}, \phi, \tilde{\rho}, \tilde{P}, \tilde{u}^b \rangle \quad (+) \quad \tilde{\nabla}_a(\tilde{\rho}\tilde{u}^a) + \tilde{P}\,\tilde{\nabla}_a \tilde{u}^a = -\phi^{-1}(\tilde{\rho} - 3\tilde{P})\tilde{u}^a \tilde{\nabla}_a \phi,$$
$$(\tilde{\rho} + \tilde{P})\tilde{u}^b \tilde{\nabla}_b \tilde{u}^a + \tilde{h}^{ab}\tilde{\nabla}_b \tilde{P} = (\tilde{\rho} - 3\tilde{P})\tilde{h}^{ab}\tilde{\nabla}_b \log \phi, \tag{PF'}$$

where $\tilde{\nabla}$ is the derivative operator with respect to $\tilde{g}_{ab} = \phi^2 \eta_{ab}$, $\tilde{h}^{ab} = \tilde{g}^{ab} + \tilde{u}^a \tilde{u}^b$, $\tilde{u}^a = \phi^{-1} u^a$, $\tilde{\rho} = \phi^{-4}\rho$ and $\tilde{P} = \phi^{-4}P$. (See Appendix A for further details.) However, this formulation introduces the extra field $\phi$, which is dispensable from the perspective of our original formulation of the theory.[7] Note that in (MAXW'), in contrast with (PF'), $\phi$ is not present in the dynamical equations. We will refer to this fact as a conformal symmetry, which will become important for Section 3.

---

4 Fields are smooth maps that assign a field value $\psi_i(x)$ to each point $x \in U$, where $U \subseteq M$ is an open set.

5 Our convention in the rest of the paper is $c = 1$ and $8\pi G = 1$.

6 This excludes, for instance, Palatini-type approaches to gravitational theories. For a review, see Olmo (2011); for a philosophical discussion, see Duerr (2021).

That relativistic field theories admit multiple representations is, in itself, unproblematic for analyzing the STM dichotomy in the examples above. However, this feature will present significant challenges in the context of scalar–tensor theories. The remainder of this section focuses on addressing the STM dichotomy within relativistic field theories, with the aim of formulating a *rigorous* classification scheme.

By *rigorous*, we mean that for any given representation of a relativistic field theory (RFT), the procedure assigns each element unambiguously to the spacetime or the matter class. This framework will then be applied to test whether such a disjoint classification remains tenable in the case of scalar–tensor theories.

## 2.2 SPACETIME AND MATTER CRITERIA FOR RELATIVISTIC MODELS

To classify the elements of a relativistic field theory as either matter or spacetime, we adopt two distinct (but interdependent, as we will see) criteria grounded in recent philosophical analysis. For matter, we follow the approach developed in Martens and Lehmkuhl (2020a), who identify three key properties, in order of increasing strength, that characterize matter fields in relativistic theories: (i) they exhibit evolving degrees of freedom (as described by differential equations), (ii) they can mediate interactions with other fields (in a symmetric way, i.e., when one field acts upon another, the other reacts), and (iii) they are capable of carrying and transferring energy-momentum. These properties provide a robust and empirically grounded basis for identifying matter fields across different theories and their representations. In contrast, our criterion for spacetime has a narrower scope, since our analysis is restricted to metric field theories.[8] We consider as spacetime only those fields that determine the geometric background—namely, the (possibly composite) Lorentzian metric $g_{ab}$ from which the covariant derivative operator is constructed via the compatibility condition $\nabla g = 0$. This suffices to ensure the usual properties associated with spacetime, such as inertial and causal structure.

---

7 This apparent dispensability might lead one to argue that such a representation is artificial compared to (PF). This critique has been defended in Duerr (2021) regarding $F(R)$ gravity theories, where both $\langle M, g_{ab} \rangle$ and $\langle M, g_{ab}, \phi \rangle$ serve as possible but unequally parsimonious theories.

8 For a broader discussion of defining spacetime, see, e.g., Knox (2019); Baker (2021).

With these criteria, we aim to develop a precise and consistent method for assigning each field in a (representation of a) relativistic field theory to either the spacetime or matter sector. This classification will then be tested in the context of scalar–tensor theories, where the distinction becomes more subtle and representation-dependent.

The present criteria are intended to be minimal in the following sense. The central feature underlying the STM distinction is the conservation of the stress-energy tensor. The existence of a stress-energy tensor is the condition associated with matter fields, whereas the conservation holds with respect to the derivative operator constructed out of the spacetime fields. The remaining conditions formalize functional features associated with stress-energy tensors, such as their role in dynamical interaction and energy exchange. Guided by a principle of interpretative charity, the criteria are constructed so as to recover the standard STM dichotomy in paradigmatic relativistic theories, especially GR coupled to ordinary matter fields. Thus, the argument does not rely on highly revisionary or theory-specific definitions, but rather on a minimal formalization of distinctions already implicit in relativistic field theory.

### 2.2.1 Matter criterion for relativistic models

In relativistic field theories, matter consists of fields. Matter may consist of a single field (e.g., the electromagnetic field (MAXW)) or of several fields (e.g., fluids (PF)). These fields are required to have well-defined dynamics, encoded in the dynamical equations of the RFT. To make this idea rigorous, we follow results from Geroch (1996). Given the set of dynamical equations for a particular RFT, we can associate a subset of those equations with a particular (matter) field as follows.

**Definition 1 (Dynamical equation of a field)** Let $\psi$ be a field that is part of a system of partial differential equations. Write the system in first-order form, introducing auxiliary fields if needed, so that higher-order derivatives are eliminated, e.g. $\nabla_a \nabla_b \psi =: \nabla_a \Psi_b$. If $\nabla\Psi$ appears in an equation of the system, then that equation is a *dynamical equation* for $\psi$ (see Appendix B for a more formal discussion). If a field has at least one dynamical equation, we call it a *dynamical field*.

With this definition, the condition for matter to exhibit dynamics translates into the requirement that each field in our matter class possesses at least one dynamical equation, in the sense of Def. 1. Applied to the example of Maxwellian electromagnetism (MAXW), we can clearly conclude that $A_a$ satisfies this aspect of being a matter field. This is not the case for $j^b$: for a given electromagnetic current, the standard assumption is that there exists some underlying matter field(s) from which $j^b$ can be derived. For instance, in a joint system of electromagnetism and a (charged) perfect fluid (Hawking and Ellis 1973):

$$\begin{aligned} \langle M, \eta_{ab}, A_a, \rho, P, u^a \rangle \quad (+) \quad & \nabla_a F^{ab} = eu^b, \quad \nabla_{[a} F_{bc]} = 0, \\ & (\nabla_a \rho) u^a + (\rho + P) \nabla_a u^a = 0, \\ & (\rho + P) u^b \nabla_b u^a = -h^{ab} \nabla_b P + e F^a_b u^b. \end{aligned} \qquad \text{(MAXW-PF)}$$

In (MAXW-PF), instead of $\langle A_a, j^b \rangle$ being the set of dynamical fields, we have $\langle A_a, \rho, P, u^a \rangle$. The difference between (MAXW) and (MAXW-PF) is important: in the former, the electromagnetic source appears in the dynamical equation for $A_a$ but not vice versa, because the source is given externally. In the latter, both $A_a$ and $\langle \rho, P, u^a \rangle$ appear in each other's dynamical equations. It is for this reason that $j^b$ is not matter, but $\langle \rho, P, u^a \rangle$ will be. Inspired by the nomenclature of Lehmkuhl (2010), we propose:

**Definition 2 (Coupling/ Interaction)** Two fields $\psi_1$ and $\psi_2$ are *directly coupled* if each field appears in at least one of the dynamical equations of the other. We say they are *directly un-coupled* if neither of them appears in the other's dynamical equation. An *interacting system* is a collection of dynamical fields such that every field is directly coupled to at least one other field.

This definition of coupling recovers the familiar aspect of matter as that which acts upon, and is acted upon by, other matter. Therefore, we require any matter field to either be directly coupled or directly un-coupled to any other matter field. In (MAXW-PF), $A_a$ and $\langle \rho, P, u^a \rangle$ are dynamical fields coupled directly. The metric $\eta_{ab}$ is neither directly coupled nor directly un-coupled to $A_a$ and $\langle \rho, P, u^a \rangle$ since it is present in their dynamical equations, but not the other way round.[9] This motivates the following necessary criterion for being matter:

> **Criterion 1 (Action–Reaction)** For a given relativistic field theory, and for the set of DPMs in a particular representation $\langle M, \psi_1, ..., \psi_n \rangle$, we classify a subset $\langle \psi_i, ..., \psi_j \rangle$ as matter only if:
> (i) all of its members are dynamical fields (Def. 1), and
> (ii) each field is either directly coupled or directly un-coupled to any other field (see Def. 2).

The notion of interacting system allows us to generalize the concept of interaction between fields to cases where two fields are directly un-coupled but, for instance, share an intermediate field that is directly coupled to both of them (which in Lehmkuhl (2010) is called indirect coupling). This will become necessary for the fuller concept of matter adopted below.

A further condition that is typically insisted upon concerns the notion of energy. The existence of a stress-energy tensor is often taken as a requirement for relativistic matter fields (see, e.g., sec. 1.2 of Wald (2022) for the electromagnetic case). Moreover, Fletcher (2025) argues that matter differs from spacetime precisely in this respect:

> Matter fields are just those fields for which there is an explicit procedure specifying how their variables contribute to the energy-momentum. [...] That matter fields interact means that they have the potential to exchange energy and momentum, leading to differences in their dynamics. [...] Fields that represent spatiotemporal structure are just those that do not contribute variably to the energy-momentum.[10]

Accordingly, we require that any material system composed of matter fields $\psi_i$ possesses a stress-energy tensor $T^{ab}$. For instance, for electromagnetism:

$$T^{ab}_{\text{EM}} = F^{ac} F^{b}_{c} - \frac{1}{4} g^{ab} F_{cd} F^{cd}.$$

Furthermore, the stress-energy tensor of the full system must be covariantly conserved with respect to the derivative operator appearing in the dynamical equations:

$$\nabla_a T^{ab} = 0.$$

9 Note that although in the representation (MAXW-PF) we have not explicitly shown it, we could add the extra equation $R^{d}{}_{abc}[\eta] = 0$ as a dynamical equation (Pooley 2017), but the same conclusion would hold.

10 On Fletcher's definition, matter and spacetime are mutually exclusive, thereby insisting upon a black-and-white STM dichotomy *by fiat* (apart from perhaps allowing fields that are neither matter nor spacetime). When defining the concept of spacetime below, we see no reason to add to those positive features the constraint that spatiotemporal fields cannot have energy, and thus cannot also be matter.

The existence and conservation of $T^{ab}$ serves two purposes. First, if in addition there exists a timelike Killing vector field (e.g., in flat spacetime), there is a conserved energy–momentum for the full system. Second, if we can decompose the matter sector into interacting systems (see Def. 2), then each interacting system admits its own covariantly conserved stress-energy tensor $T_i^{ab}$ such that

$$T^{ab} = \sum_i T_i^{ab}, \quad \nabla_a T_i^{ab} = 0.$$

In some cases, a subsystem's $T_i^{ab}$ can be further decomposed into stress-energy tensors representing distinct matter components within that subsystem, allowing us to formulate interaction as an exchange of energy–momentum between components. For example, for an electromagnetic field interacting with a perfect fluid in flat spacetime (MAXW-PF), the associated stress-energy tensor (Hawking and Ellis 1973) is

$$T^{ab} = \underbrace{\varrho u^a u^b + P h^{ab}}_{T^{ab}_{\mathrm{PF}}} + \underbrace{F^{ac} F^b_{\ c} - \frac{1}{4} g^{ab} F_{cd} F^{cd}}_{T^{ab}_{\mathrm{EM}}},$$

which is conserved, and hence $\nabla_a T^{ab}_{\mathrm{PF}} = -\nabla_a T^{ab}_{\mathrm{EM}}$, expressing precisely the transfer of energy–momentum between the fluid and the field. All of this motivates the following complete criterion for matter:

> **Criterion 2 (Matter)** For a given relativistic field theory, and for the family of DPMs in a particular representation $\langle M, \psi_1, ..., \psi_n \rangle$, we classify a subset $\langle \psi_i, ..., \psi_j \rangle$ as matter iff:
> (i) it satisfies the action-reaction principle (Criterion 1);
> (ii) for any given DPM, there exists a stress-energy tensor $T^{ab}$ satisfying $\nabla_a T^{ab} = 0$ with respect to a derivative operator $\nabla$; and
> (iii) the stress-energy tensor $T^{ab}$ can be further decomposed into covariantly conserved stress-energy tensors $T^{ab} = \sum_i T_i^{ab}$ with $\nabla_a T_i^{ab} = 0$ iff the matter system can be divided into interacting subsystems each of them having assigned one stress-energy tensor $T_i^{ab}$.

We illustrate this criterion by checking whether it gives the expected verdict when applied to our earlier examples: $A_a$ is indeed part of the matter class, while $j^b$ is not (as $j^b$ can be derived from a charged fluid, itself a matter field). The perfect fluid is also a matter field. In both examples, $\eta_{ab}$ cannot be part of the matter class, since it does not satisfy either criterion. All is as expected.

Finally, consider whether $g_{ab}$ in (EFE) can belong to the matter category. The answer is no: it satisfies the action–reaction principle (Crit. 1) but not the full matter criterion (Crit. 2). To see this, suppose the only non-vanishing field in the set of matter fields is $g_{ab}$ (non-vanishing by construction). Condition (ii) of the matter criterion would then require a conserved stress-energy tensor constructed from $g_{ab}$ alone, which is either impossible under mild conditions (Curiel 2019) or, at best, arbitrary (Fletcher 2025). Furthermore, even in the presence of another matter field, e.g., an electromagnetic field, $A_a$ and $g_{ab}$ are directly coupled, but nevertheless there is a conserved stress-energy tensor associated with the former, which would be contrary to condition (iii) of the matter criterion (Crit. 2).

### 2.2.2 Spacetime criterion for relativistic models

Standard textbooks associate spacetime with a manifold plus additional tensor fields. For example, Newtonian spacetime is often formulated in terms of $\langle M, h^{ab}, t_{ab}, \nabla^a \rangle$ and derived notions, and special relativity in terms of $\langle M, \eta_{ab} \rangle$ (see Earman (1989); Read (2023) for detailed accounts). Such structures provide precise definitions of causal, inertial, and clock structures (Malament 2012; Weatherall 2016). In relativistic models, the Lorentzian metric $g_{ab}$ defines straight curves and classifies tangent vectors as timelike, null, or spacelike. Ehlers writes:

> Spacetime is a smooth 4-manifold $M$ with a pseudo-Riemannian metric $g_{ab}$ of signature $(+ - --)$. Its null geodesics represent light rays, the timelike geodesics of the Riemannian connection $\Gamma^a_{bc}$ associated with $g_{ab}$ represent the worldlines of freely falling test particles, and the arc length along timelike lines measures the proper time shown by a standard clock. (Ehlers 1973, p.82).

Note that causal and inertial structures are here defined via test particles and light rays. One approach to defining spacetime is to start with primitive notions of these two tracers of spacetime and reconstruct the geometric structure—this is the Ehlers–Pirani–Schild program (Adlam, Linnemann and Read 2025; Ehlers, Pirani and Schild 2012). While attractive, it would require treating test particles as matter, which is problematic: their stress-energy tensor does not, by definition, appear in Einstein's equations (Fletcher 2025). Since we are working with field theories, we instead adopt the *metric postulate*:

> (i) that gravity is associated, at least in part, with a symmetric tensor field, the "metric"; and (ii) that the response of matter and fields to gravity is described by $\nabla T = 0$, where $\nabla$ is the divergence with respect to the metric, and $T$ is the stress-energy tensor for all matter and nongravitational fields. (Thorne and Will 1971, p.597)

This postulate was already assumed in the matter criterion (Crit. 2), but is stated here to highlight that the derivative operator $\nabla_a$ must be derived from some $g_{ab}$. We propose then the following criterion for spacetime.[11]

> **Criterion 3 (Spacetime)** Given a representation of a relativistic theory in which matter fields have been identified via criterion 2, i.e.,
>
> $$\langle M, \psi_1, ..., \underbrace{\psi_i, ..., \psi_n}_{\text{Matter}} \rangle \quad + \quad \text{Dynamical equations} \quad + \quad \text{Stress-energy tensor for matter}\,,$$
>
> spacetime consists of: (i) the manifold $M$, (ii) the metric $g_{ab}$ used to define the derivative operator satisfying $\nabla_a T^{ab} = 0$, and (iii) all fields of the DPM, if any, used to construct $g_{ab}$.

The proposed criterion rests on a minimal yet physically meaningful foundation: it identifies as spacetime precisely those elements of a relativistic theory needed to construct $g_{ab}$, which in turn defines the covariant derivative $\nabla_a$ satisfying $\nabla_a T^{ab} = 0$. This reflects the central structural role of the metric in governing the dynamics and conservation of matter fields.[12] While idealizations such as test particles or ray optics can illustrate the geometric content of $g_{ab}$—notably its link to timelike and null geodesics—these limits are themselves idealized and may not be derivable from arbitrary matter configurations.[13] By contrast, the requirement that $g_{ab}$ underwrite a derivative operator with conserved matter stress-energy applies universally across well-posed relativistic field theories, independent of the specific dynamics or any idealizations applied to the matter content. Thus, the spacetime category must include the manifold $M$ and all DPM fields necessary to construct $g_{ab}$, which provides the geometrical structure for matter dynamics.

---

11 In Martens and Lehmkuhl (2020a), an additional criterion is insisted upon, the philosopher's Strong Equivalence Principle: special relativity should be locally valid, i.e., an absence of curvature terms in the equations of motion to ensure that rods and clocks do locally behave as in special relativity. As we will see in footnote 17, imposing this further condition would make things even worse for the STM dichotomy..

12 One may ask why $g_{ab}$ is given this privileged role, rather than an independently specified connection, as in Palatini or metric-affine formulations. This is a fair question, but it lies outside the scope of the present case study. Here we restrict attention to metric formulations of relativistic field theories in which the relevant derivative operator is the torsion-free, metric-compatible Levi-Civita connection determined by $g_{ab}$. Our criterion is therefore not intended to exclude broader notions of spacetime structure in theories with independent connections, torsion, or non-metricity; such cases require a separate analysis.

We have now introduced criteria for matter and spacetime. For the STM dichotomy to hold, each element that goes into the DPMs of a theory must be classified as either spacetime or matter:

$$\langle \underbrace{M, \psi_1, \dots,}_{\text{Spacetime}} \underbrace{\psi_i, \dots, \psi_n}_{\text{Matter}} \rangle \quad + \quad \text{Dynamical equations} \quad + \quad \text{Stress-energy tensor for matter}\,.$$

Figure 1 summarizes the steps for testing whether a given (representation of a) relativistic field theory upholds the dichotomy.

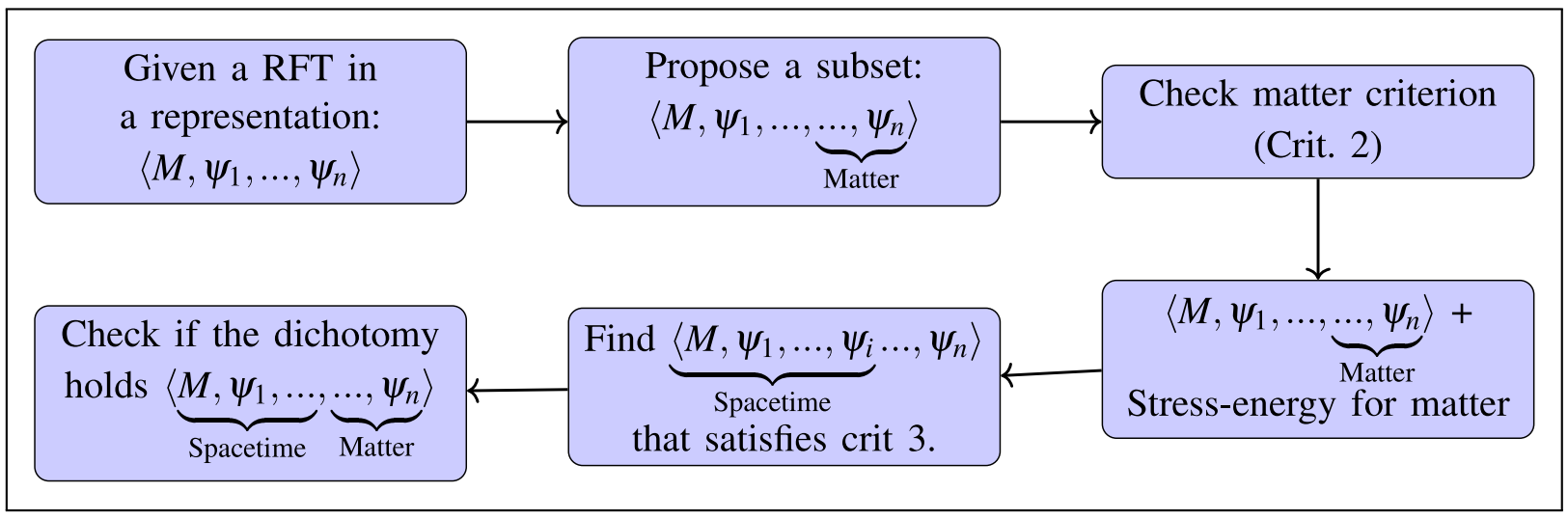

**Figure 1** Different phases for testing the spacetime matter dichotomy for a given model.

The crux here is that one cannot categorize a single field in isolation. A corollary of this is that, if one adds a new field to a theory, say to GR, then one needs to reverify the categorization for all fields. The reason the categorization has a collective nature is twofold. First, one can only determine of a whole group of matter candidates whether they are indeed matter because for each pair of them the action–reaction principle needs to be satisfied (Criterion 1). Second, each new field could contribute to the tensor that defines the derivative operator with respect to which the stress-energy tensor for matter is conserved (Criterion 3.ii).

Let us now see how the dichotomy applies to two paradigmatic examples: a perfect fluid in an electromagnetic field, and a perfect fluid in a gravitational field, both of which we then consider in the pressureless limit.

| Perfect fluid in an electromagnetic field | Perfect fluid in a gravitational field |
|---|---|
| $(\rho + P)u^b \nabla_b u^a = -h^{ab}\nabla_b P + eF^a_b u^b$ | $(\rho + P)u^b \nabla_b u^a = -h^{ab}\nabla_b P$ |
| $\nabla_a F^{ab} = eu^b, \quad \nabla_{[a}F_{bc]} = 0$ | $R^{ab} - \frac{1}{2}Rg^{ab} = T^{ab}$ |
| $T^{ab} = \underbrace{\rho u^a u^b + Ph^{ab}}_{T^{ab}_{\text{PF}}} + \underbrace{F^{ac}F^b_c - \frac{1}{4}g^{ab}F_{cd}F^{cd}}_{T^{ab}_{\text{EM}}}$ | $T^{ab} = \underbrace{\rho u^a u^b + Ph^{ab}}_{T^{ab}_{\text{PF}}}$ |
| $\langle \underbrace{M, \eta_{ab},}_{\text{Spacetime}} \underbrace{A_a, \rho, P, u^b}_{\text{Matter}} \rangle$ | $\langle \underbrace{M, g_{ab},}_{\text{Spacetime}} \underbrace{\rho, P, u^b}_{\text{Matter}} \rangle$ |

From the stepwise process in Figure 1, we conclude that the dichotomy holds in both cases. The key difference between the two is that, in the electromagnetic case, there exists a

13 More sophisticated methods can recover timelike and null geodesics from $\nabla_a$ by imposing conservation plus an energy condition (Geroch and Weatherall 2018; Weatherall 2020), but such results require particular types of solutions, not guaranteed for all DPMs.

stress-energy tensor for the electromagnetic field, and conservation of the total stress-energy tensor can be shown to yield—in the limit $P \to 0$:

$$\nabla_a T^{ab} = 0 \quad \Rightarrow \quad u^a \nabla_a u^b = \frac{e}{\rho}\, u^c F_c{}^b,$$

which is the geodesic equation with respect to $\eta_{ab}$, modified by the Lorentz force from the electromagnetic field. In the gravitational case:

$$\nabla_a T^{ab} = 0 \quad \Rightarrow \quad u^a \nabla_a u^b = 0,$$

which is precisely the geodesic equation with respect to $g_{ab}$ (see Appendix A for more details). Thus, in both cases, the geometric interpretation of spacetime can be recovered from an appropriate limit[14] (e.g., a pressureless perfect fluid), and this is fully consistent with our classification scheme: spacetime provides the background geometry relative to which the matter fields' motion is determined, while matter fields carry energy–momentum which they transfer if interacting.

## 3 STM DICHOTOMY FOR SCALAR–TENSOR THEORIES

In this section, we turn to scalar–tensor theories and examine how the STM classification fares in this more complex setting. Scalar–tensor models are distinctive in that, besides the metric $g_{ab}$ and matter fields, they contain an additional scalar field whose dynamics are intertwined with those of the metric. We find that this "mixing" complicates the task of classification. A central question will be whether the scalar field can be consistently regarded as matter or as part of spacetime, and how this status depends on the chosen representation of the theory.

To address this, we will analyze the Jordan frame and two Einstein-frame representations. In the Jordan frame, the scalar field $\phi$ cannot be classified as either matter or spacetime: it violates the matter criteria because of one-way couplings to other fields, and it fails to satisfy the spacetime criteria because the conservation law involves only the metric $g_{ab}$. By contrast, in the Einstein frame, we encounter two distinct possibilities. In one representation, the redefined scalar $\theta$ must be included in the spacetime sector together with the new metric $\tilde{g}_{ab}$; in the other, $\theta$ can be consistently classified as matter but not as spacetime. Thus, depending on the representation, the scalar field oscillates between these categories, implying a breakdown of the STM dichotomy.

We will conclude in Section 3.3 by exploring the consequences of this representational dependence. First, the geodesic motion of fluids confirms the internal consistency of each classification. Second, we observe that across all frames, matter fields couple to the metric–scalar pair only through the stress-energy tensor. This leads us to propose the notion of "gravitational field," encompassing those fields whose dynamical equations involve matter exclusively through the stress–energy tensor. Unlike the STM dichotomy, this classification is frame-independent. As such, it offers a more robust way of capturing the role played by the metric and scalar fields in scalar–tensor theories.

14 In the case of a fluid, one can see already from the dynamical equations that the geodesic limit (or deviation due to the Lorentz force) will appear in the limit of $P \to 0$. Our point is that the force appears more generally as an interchange of stress-energy mediated by the conservation law $\nabla_a T^{ab} = 0$.

### 3.1 SCALAR–TENSOR THEORIES

Let us first define a scalar–tensor theory.

**Definition 3 (Scalar–tensor theory)** A scalar–tensor theory has in at least one representation the form

$$\langle M, g_{ab}, \phi, \psi_1, \ldots, \psi_n \rangle$$

where $\psi_i$ are some matter fields and the dynamical equations for the scalar field $\phi$ and metric field $g_{ab}$ are

$$F(\phi)G^{ab} - G(\phi)(\nabla^a\phi\nabla^b\phi - \frac{1}{2}g^{ab}\nabla^c\phi\nabla_c\phi) - \nabla^a\nabla^b F(\phi) + g^{ab}\Box F(\phi) + \frac{1}{2}V(\phi)g^{ab} = T_M^{ab}[\psi_i, g],$$
$$2G(\phi)\Box\phi + G'(\phi)\nabla^c\phi\nabla_c\phi - V'(\phi) + F'(\phi)R = 0, \tag{JF}$$

with the conditions i) $\nabla_a T_M^{ab}[\psi_i, g] = 0$, ii) $F(\phi) > 0$, iii) $2F(\phi)G(\phi) + 3F'(\phi)^2 > 0$ (which guarantees that the field redefinition to the Einstein frame below is well defined), and iv) the stress-energy tensor and the dynamical equations for the matter fields $\psi_i$ depend only on the matter fields $\psi_i$ and the metric $g_{ab}$, but not on $\phi$.[15]

We call this representation of the theory the Jordan frame.[16] This representation has two key features.[17] First, the presence of the scalar field multiplying the Einstein tensor on the left-hand side, which mimics an effective Newton's constant. Second, the existence of a *mixing problem* (Fujii and Maeda 2003), namely the appearance of second-derivative terms ($\nabla\nabla\phi$) in both equations. A way of overcoming the mixing problem is changing to a different (more suitable) representation, known in the literature as a *frame transformation*. This change of representation involves a metric (conformal) transformation $g_{ab} \to \tilde{g}_{ab}$ and a scalar field transformation $\phi \to \theta$ (see Appendix C). In this new representation, *the Einstein frame*, the equations of the scalar–tensor theory are

$$\tilde{G}^{ab} - \tilde{\nabla}^a\theta\tilde{\nabla}^b\theta + \frac{1}{2}\tilde{g}^{ab}\tilde{g}^{cd}\tilde{\nabla}_c\theta\tilde{\nabla}_d\theta + U(\theta)\tilde{g}^{ab} = \frac{1}{F^3(\theta)}T_M^{ab}[\psi_i, F^{-1}\tilde{g}].$$
$$\tilde{\Box}\theta - U'(\theta) = \frac{F'}{2F^4}\tilde{g}_{ab}T_M^{ab}[\psi_i, F^{-1}\tilde{g}] \qquad \nabla_a T_M^{ab}[\psi_i, F^{-1}\tilde{g}] = 0, \tag{EF}$$

where now the two fields $\tilde{g}_{ab}$ and $\theta$ are present and $U = V/2F^2$; from here on, primes on functions of $\theta$ denote derivative with respect to $\theta$. Note that the conservation of the stress-energy tensor is still with respect to the derivative operator constructed out of $g_{ab}$. In order to write a conservation law with respect to the new metric, it is convenient to further transform the stress-energy tensor of matter as follows

$$\tilde{T}_M^{ab}[\psi_i, \theta, \tilde{g}] = F^{-3}(\theta)T_M^{ab}[\psi_i, F^{-1}\tilde{g}]$$

The scalar–tensor equations (EF) then become (see Appendix C)

$$\tilde{G}^{ab} = \tilde{T}_M^{ab}[\tilde{\psi}_i, \tilde{g}, \theta] + T_\theta^{ab}[\theta, \tilde{g}],$$
$$\tilde{\Box}\theta - U'(\theta) = \frac{F'}{2F}\tilde{g}_{ab}\tilde{T}_M^{ab}[\tilde{\psi}_i, \tilde{g}, \theta], \qquad \tilde{\nabla}_a\left(\tilde{T}_M^{ab}[\psi_i, \tilde{g}, \theta] + T_\theta^{ab}[\theta, \tilde{g}]\right) = 0, \tag{EF'}$$

15 Note that the scalar field $\phi$ is not *reducible* to the metric $g_{ab}$ in the sense discussed by Duerr (Duerr 2021) in the context of the Brans–Dicke theory (a sub-type of scalar tensor theory) constructed out of a given $F(R)$ theory.

16 Note that this has nothing to do with the notion of reference frame.

17 A third interesting feature is that there is a term with the curvature scalar. This means that the (philosopher's) Strong Equivalence Principle (fn. 11) is not satisfied. If this were taken on board as an additional spacetime criterion, the metric in the Jordan representation would not even be spacetime, wreaking further havoc for the STM dichotomy..

where $T_{\theta}^{ab}[\theta, \tilde{g}] = \tilde{\nabla}^a\theta\tilde{\nabla}^b\theta - \frac{1}{2}\tilde{g}^{ab}\tilde{g}^{cd}\tilde{\nabla}_c\theta\tilde{\nabla}_d\theta - U(\theta)\tilde{g}^{ab}$. Finally, we also transform each matter field

$$\psi_i \to \tilde{\psi}_i = F(\theta)^{p_i}\psi_i,$$

where $p_i$ is a field-dependent constant. Note that this transformation can also be done for the frame (EF). The behavior of the dynamical equations for each field lets us classify them as follows.

**Definition 4 (Conformally invariant fields)** We say a field $\psi_i$ is conformally invariant if one can find values for the $p_i$'s such that the transformations $g_{ab} \to \tilde{g}_{ab}$ and $\psi_i \to \tilde{\psi}_i$ leave the dynamical equations invariant. Furthermore, we require that the stress-energy tensor associated with an interacting system consisting only of conformally invariant fields has vanishing trace.[18] If either condition is not satisfied, we say $\psi_i$ is a non-conformally invariant field or that the interacting matter system is not conformally invariant .

For instance, the electromagnetic field is a conformally invariant field. This can be seen by comparing (MAXW) and (MAXW') and taking the trace of the stress-energy tensor. For an interacting system consisting of only conformally invariant fields, we have

$$\nabla_a T^{ab} = 0 \implies \tilde{\nabla}_a \tilde{T}^{ab} = 0,$$

whereas in the case of an interacting system consisting of non-conformally invariant fields, e.g., (PF) and (PF'), we have

$$\nabla_a T^{ab} = 0 \implies \tilde{\nabla}_a \tilde{T}^{ab} = -\tilde{g}^{ab}\tilde{T}^{cd}\tilde{g}_{cd}\tilde{\nabla}_a \log F^{1/2}.$$

See Appendix D of Wald (1984) for the proof.

## 3.2 STM DICHOTOMY

We now evaluate the STM dichotomy for scalar–tensor theories. As we will see along the way, all fields except $\phi$ (or $\theta$) are correctly classified: the metric $g_{ab}$ (and $\tilde{g}_{ab}$) are part of spacetime, and the rest of the fields are part of matter.[19] The main question: how to classify the scalar field in each frame ($\phi$ or $\theta$)?

### 3.2.1 Matter criterion

Let us start by classifying $\phi$:

$$\langle \underbrace{M, g_{ab}}_{\text{ST}}, \underbrace{\phi}_{?}, \underbrace{\psi_1, \ldots, \psi_n}_{\text{Matter}} \rangle. \qquad \text{(STM-Jordan)}$$

The first step is to check whether $\phi$ satisfies the matter criterion. It does not.

**Conclusion 1 (Jordan Frame – Matter)** The scalar field $\phi$ (in the Jordan Frame) cannot be consistently classified as matter according to the matter criterion (Criterion 2).

The proof is straightforward once we note that, according to the definition of scalar–tensor theories, the field $\phi$ does not appear in the dynamical equations of the matter fields $\psi_i$ (i.e., it does not react to any action it might get from those fields). However, the first equation of (JF) is a dynamical equation for both the metric $g_{ab}$ and $\phi$ (see Appendix B). Therefore, all types of matter appear in the dynamical equations of $\phi$ (i.e., they do indeed act on $\phi$). In other words, $\phi$ is neither directly coupled nor directly un-coupled to the matter fields.

---

18 We note that this condition follows directly if we compute the stress-energy tensor via an action (see Wald (1984)).

19 As discussed in footnotes 11 and 17, if we further impose the Strong Equivalence Principle when defining spacetime, the metric $g_{ab}$ in the Jordan representation would not be spacetime either..

Hence, $\phi$ does not satisfy the action–reaction principle (Criterion 1) and *a fortiori* also does not satisfy the matter criterion (Criterion 2).

This behavior of scalar–tensor theories, i.e., the one-way coupling of matter fields to $\phi$, has its origin in the *mixing problem* and therefore it is wise to analyze the unmixed Einstein frames. In this representation, we have the following models

$$\langle \underbrace{M, \tilde{g}_{ab}}_{\text{ST}}, \underbrace{\theta}_{?}, \underbrace{\psi_1, \ldots, \psi_n}_{\text{Matter}} \rangle. \qquad \text{(STM-Einstein)}$$

We want to determine whether $\theta$ can be considered a matter field. Indeed, with $\theta$ added as candidate matter field to the other matter fields ($\psi_i$), this collection of fields satisfies the action–reaction principle and also the full matter criterion (Criterion 2). This is easy to check. We first consider the coupling of $\theta$ in the dynamical equations of the system in (EF). Only fields with a stress-energy tensor with a non-vanishing trace appear in the dynamical equations of $\theta$. In turn, $\theta$ also appears in their dynamical equations since these fields are non-conformally invariant (by Definition 4). Conformally invariant fields, such as the electromagnetic field, do not appear in the dynamical equations of $\theta$ and neither does $\theta$ appear in their dynamical equations. So, $\theta$ and non-conformally invariant fields are directly coupled, and $\theta$ and conformally invariant fields are directly un-coupled. The action–reaction principle is obeyed (Criterion 1). However, the full matter criterion is not obeyed in (EF) because there is no conserved stress-energy tensor associated with $\theta$. However, by using not (EF) but (EF') as a representation, there does exist a stress-energy tensor that is conserved. Finally, if there were a matter system formed only out of conformally invariant fields, the stress-energy tensor could be further split into one assigned to $\theta$ and another one assigned to that system, both of them independently conserved. As such, criterion 2 is satisfied.

**Conclusion 2 (Einstein Frames – Matter)** The scalar field $\theta$ can be consistently classified as part of the matter category in the representation (EF'), but not in the representation (EF).

### 3.2.2 Spacetime criterion

Having determined whether the scalar field is a matter field in each of the different representations, we now consider the spacetime criterion (Criterion 3). In the Jordan frame (JF), the conservation law for the stress-energy tensor of matter (which does not include $\phi$) is $\nabla_a T_M^{ab} = 0$, where $\nabla_a$ is constructed out of (only) $g_{ab}$, which is a spacetime field. Since $g_{ab}$ is the only field we need, $\phi$ is not part of spacetime.

**Conclusion 3 (Jordan Frame – Spacetime)** The scalar field $\phi$ (in the Jordan Frame) cannot be consistently classified as spacetime according to the spacetime criterion (Criterion 3).

In the Einstein Frame (EF), $\theta$ is also not part of the matter category, and therefore the same holds true, i.e., $\nabla_a T_M^{ab} = 0$, but now the metric that is needed to derive $\nabla_a$ is derived from the fields of our model $\tilde{g}_{ab}$ and $\theta$, i.e., $g_{ab} = F(\theta)^{-1}\tilde{g}_{ab}$, and therefore by our spacetime criterion both fields, $\theta$ and $\tilde{g}_{ab}$ have to be considered part of the spacetime category. Finally, in the case of (EF'), since $\theta$ is now part of the collection of matter fields, the full stress-energy tensor $\tilde{T}_M^{ab}[\psi_i, \tilde{g}, \theta] + T_\theta^{ab}[\theta, \tilde{g}]$ is conserved with respect to $\tilde{\nabla}_a$, which is derived from $\tilde{g}_{ab}$, which is therefore part of spacetime, from which it follows that $\theta$ is not part of spacetime.

**Conclusion 4 (Einstein Frames – Spacetime)** The scalar field $\theta$ can be consistently classified as part of the spacetime category in the representation (EF), but not in the representation (EF').

We are thus left with the following status of the STM dichotomy (see Table 1). In the Jordan frame (JF), $\phi$ is classified neither as spacetime nor as matter. In the Einstein frame (EF), $\theta$ is classified as spacetime but not as matter, whereas in the Einstein frame (EF'), $\theta$ is classified as matter but not as spacetime, i.e.

$$\langle \underbrace{M, g_{ab}}_{\text{ST}}, \phi, \underbrace{\psi_1, \ldots, \psi_n}_{\text{Matter}} \rangle; \tag{STM-Jordan}$$

$$\langle \underbrace{M, \tilde{g}_{ab}, \theta}_{\text{ST}}, \underbrace{\psi_1, \ldots, \psi_n}_{\text{Matter}} \rangle \quad \text{or} \quad \langle \underbrace{M, \tilde{g}_{ab}}_{\text{ST}}, \underbrace{\theta, \psi_1, \ldots, \psi_n}_{\text{Matter}} \rangle. \tag{STM-Einstein}$$

**Table 1** Classification of the scalar field in each frame.

| Frame | Spacetime | Matter | Gravity |
|---|---|---|---|
| JF | No | No | Yes |
| EF | Yes | No | Yes |
| EF' | No | Yes | Yes |

## 3.3 TWO CONSEQUENCES: GEOMETRICAL CONVENTIONALISM AND GRAVITATIONAL FIELDS

### 3.3.1 Geometrical conventionalism

The scalar field is classified as spacetime in (EF) and as matter in (EF'), but moving between (EF) and (EF') is just a matter of reorganizing dynamical equations. How do we make sense of this? To begin, note that in the Einstein frame (EF), one can construct a perfect fluid from any matter field (i.e., excluding $\theta$) and show that, in the pressureless limit (see Appendix A), the conservation law yields:

$$\nabla_a T^{ab} = \nabla_a(\rho u^a u^b) = 0 \implies u^a \nabla_a u^b = 0$$

which corresponds to timelike geodesic motion with respect to the metric $g_{ab} = F(\theta)^{-1}\tilde{g}_{ab}$. In contrast, in the transformed Einstein frame (EF') we find:

$$\tilde{\nabla}_a \tilde{T}^{ab} = \tilde{\nabla}_a(\tilde{\rho}\tilde{u}^a\tilde{u}^b) = -\tilde{\nabla}_a T^{ab}_{\theta} \implies \tilde{u}^a \tilde{\nabla}_a \tilde{u}^b = (\tilde{\nabla}_a \log F^{1/2}(\theta))\tilde{h}^{ab}.$$

This is the geodesic equation with respect to $\tilde{g}_{ab}$, modified by a pressure-like force term (cf. (PF)) arising from the field $\theta$.[20] Furthermore, since in this representation the scalar field $\theta$ is itself part of the matter sector, we can construct a perfect fluid composed solely of $\theta$ (setting all other matter fields to zero), which, in the pressureless limit, satisfies:

$$\tilde{\nabla}_a \tilde{T}^{ab}_{\theta} = 0 \implies \tilde{u}^a_{\theta} \tilde{\nabla}_a \tilde{u}^b_{\theta} = 0.$$

This demonstrates the internal consistency of the spacetime–matter classification and the corresponding geodesic behavior of fluids. In the Einstein frame (EF), the metric $g_{ab}$ defines the spacetime whose geodesics are traced by the (non-$\theta$) fluid. In the transformed Einstein frame (EF'), the metric $\tilde{g}_{ab}$ plays this role, but non–conformally invariant (standard) matter fields experience an additional force (mediated by $\theta$-matter) which accounts for the exchange of stress–energy between them. For a quadratic potential $U(\theta)$, the matter field $\theta$ does not interact with itself; a perfect fluid composed purely of $\theta$ thus follows geodesics with respect to $\tilde{g}_{ab}$ (in the pressureless limit).

These results may be taken as supporting a form of geometric conventionalism, the view that physical geometry is not an objective feature of the world but a matter of convention: "[c]laims about geometry result from conventional stipulations—acts of human decision making" (Dürr and Read 2024, p.3). Our aim here is not to defend or

20 Such a term can drive accelerated expansion during inflation; see Baumann (2022); Martin (2020).

critique this position, but rather to underscore the importance of the spacetime–matter dichotomy for debates concerning the conventionality of geometry. Indeed, in scalar–tensor theories, where the same field can shift between spacetime and matter roles across representations, this dichotomy shows that what counts as "geometry" is not fixed but frame-dependent—precisely the kind of ambiguity that fuels conventionalist arguments. This conclusion is reinforced when we consider the case of GR (EFE). By performing a conformal transformation analogous to the one used in the scalar–tensor analysis ($g_{ab} \to \tilde{g}_{ab} = \Omega^2 g_{ab}$, see Section 3), we obtain:

$$G_{ab} = \tilde{G}_{ab} + 2\tilde{\nabla}_a\tilde{\nabla}_b\gamma - 2\tilde{g}_{ab}\tilde{\nabla}_c\tilde{\nabla}^c\gamma + 2\tilde{\nabla}_a\gamma\tilde{\nabla}_b\gamma + \tilde{g}_{ab}\tilde{\nabla}_c\gamma\tilde{\nabla}^c\gamma = T^M_{ab}, \qquad \gamma := \log\Omega.$$

As in the scalar–tensor case, this transformation introduces a force-like term $f_a = \nabla_a\gamma$ (see Appendix A). One might take this as further evidence for geometric conventionalism in GR. However, a closer examination of the STM dichotomy challenges this interpretation. According to our Criteria 1–3, the field $\gamma$ (or equivalently $\Omega$) is not a matter field—similar to the scalar in the Jordan frame—but qualifies as a spacetime field. Consequently, the force term in the geodesic equation is not a physical force sourced by matter; it is merely a bookkeeping effect of the conformal redefinition.[21] In other words, although GR can be written in different (conformally related) frames, they describe the same spacetime: the geometric structure remains invariant. Thus, applying frame transformations to GR does not supply arguments in favor of conventionalism.

### 3.3.2 Gravitational fields

The original scalar–tensor proposal, the Jordan–Brans–Dicke theory, sought to replace Newton's constant with a dynamical scalar field (Brans 2005). In this theory, the scalar field determines the strength of the gravitational interaction independently of the spacetime geometry measured by matter (via test particles). This feature extends to generalized scalar–tensor theories: in the Jordan frame, the scalar field modulates the effective strength of gravity but does not alter the underlying spacetime structure. In this sense, the Jordan-frame scalar field is better understood as modifying the gravitational sector rather than the spacetime structure itself. This observation motivates a broader characterization of gravitational fields, one that applies equally to both the metric and scalar fields in the Jordan frame. We therefore propose the following definition:

**Definition 5 (Gravitational Field)** A field $\psi$ is a gravitational field iff matter fields appear in its dynamical equations only through the existence of a stress-energy tensor.

According to this definition, both the metric field and the scalar field in the Jordan frame qualify as gravitational fields, since matter enters their field equations exclusively through the stress-energy tensor.

An important result is that, according to this definition, the gravitational sector is invariant across different representations of scalar–tensor theory. In the Jordan frame (JF), the gravitational fields are $g_{ab}$ and $\phi$ and the matter fields enter the dynamics only through the conserved stress-energy tensor $T_M^{ab}$. In the Einstein frame (EF), the gravitational fields are $\tilde{g}_{ab}$ and $\theta$ with matter again contributing solely through the conserved stress-energy $T_M^{ab}$. Finally, in (EF'), the gravitational fields remain $\tilde{g}_{ab}$ and $\theta$, but the stress-energy tensor must now include the $\theta$ field, since it is matter. This requirement is satisfied by the conserved total stress–energy $\tilde{T}_M^{ab} + T_\theta^{ab}$. Importantly, this analysis still relies on the matter criterion

21 For an alternative argument against geometric conventionalism in GR, see Weatherall and Manchak (2014); for a defence of conventionalism as a live option, see Dürr and Ben-Menahem (2022); for further clarification of the scope and limits of the Weatherall–Manchak result, see Mulder (2026).

(Criterion 2) to determine which fields count as matter in each frame. Remarkably, although the matter sector changes from frame to frame, the set of gravitational fields—defined through this invariant sourcing structure—remains the same.

This frame-invariant property (see Table 1) points towards a categorization that differs from the STM dichotomy: gravitational and non-gravitational fields. Spacetime (together with the manifold $M$) can always be derived solely from gravitational fields, while gravitational fields themselves may also play the role of matter.

**Conclusion 5 (All Frames – Gravity)** In each of the three frames, the associated metric and scalar field are gravitational fields (Def. 5).

## 4 PHILOSOPHICAL INTERPRETATIONS

We have argued that classifying the scalar field(s) as a form of matter or as spacetime structure is not a black-and-white matter (pun intended). The reasons are twofold: (1) the classification is not invariant across frames, i.e., across the Jordan representation versus the Einstein representation; (2) even within each of these representations, the STM dichotomy faces a problem, albeit a different problem in each case: (2a) in the Jordan representation the scalar field is neutral, i.e., neither spacetime nor matter (plus, arguably also the metric is denied the status of spacetime, see footnotes 11, 17, and 19); (2b) although in the Einstein representation the scalar field is not neutral anymore, whether it is purely spacetime *or* purely matter depends on how one defines the stress-energy tensor (EF vs. EF'). A more positive upshot seems to be that the category of gravitational field can be consistently attributed to the scalar field in all these representations. In this section, we introduce and evaluate the various interpretational strategies available from here, and argue that they all lead to a breakdown of the conceptual dichotomy between spacetime and matter, in one way or another.

A first philosophical lesson one may wish to draw from this case study is that of functionalism about, e.g., the spacetime category.

**Interpretation 1 (Spacetime Functionalism)** All there is to being spacetime is instantiating the spacetime role.

A crucial feature of this position is that of multiple realizability: it is possible for multiple things, for instance fields, even matter fields, to play the spacetime role. For instance, $g$ could play the spacetime role by itself (in the Jordan frame),[22] or $\tilde{g}$ and $\theta$ could team up to play the spacetime role (in the Einstein frame (EF)), or $\tilde{g}$ could play that role by itself (in the Einstein frame (EF')). However, adopting functionalism does not by itself address the most interesting features of our case study, i.e., the frame-dependence of the instantiator of the spacetime and matter roles. Although functionalism is consistent with our case study, it does not give the complete story.

The second interpretational decision point may be framed by analogy with the debate between the interpretationalist versus motivationalist stance towards symmetries (Martens and Read 2021; Møller-Nielsen 2017). Note that we remain agnostic here as to whether the Jordan and Einstein frames should be treated as theoretically and/or dynamically equivalent, and whether the transformations between them are to be considered symmetries, dualities, or gauge symmetries—all we need here is the (unproven but generally expected) empirical equivalence between these representations/frames. According to (strong)[23]

22 Bracketing the caveats mentioned in earlier footnotes.

23 More on the weak vs. strong versions of these two stances later.

interpretationalism (Martens and Read 2021; Saunders 2007), symmetries *immediately* imply an anti-realist commitment towards symmetry-variant structure. On this approach, symmetries *always* constitute mere redescriptions—e.g., changes of variables—of the same physical state of affairs, even if the descriptions seem metaphysically distinct and it may be unclear what the underlying, common, metaphysically perspicuous state of affairs is that is described by each set of symmetry-related models. Those who adopt the (weak) motivational stance insist that symmetries merely *motivate* us to find a new (version of the) theory that either trades only in symmetry-invariant structure (i.e., a modification of the syntax to obtain a "reduced theory") or that allows so-called internal sophistication (i.e., a semantic modification that "forgets" certain structures) (Dewar 2019; March and Read 2025; Martens and Read 2021). *Only if* such a reduced or sophisticated theory can be found can it be concluded that we should be anti-realists about the symmetry-variant structure in the old (version of the) theory, and realists about all the structure of the new (version of the) theory (which has after all been purged of the surplus structure).

As the Einstein and Jordan frames are empirically equivalent, the transformations between them resemble symmetries and dualities in the sense that matters for the above distinctions (Le Bihan and Read 2018; Read and Møller-Nielsen 2020; Martens and Read 2021, §2.2). We may then consider whether there is already a reduced or sophisticated version of scalar–tensor theory available, e.g., a (version of the) theory that knows no such frames. If so, there is no room for the interpretationalist and motivationalist to disagree.

**Interpretation 2 (Structural Common Core)** Only commit, ontologically, to the structural common core of the various frames, i.e., the physical structure that is invariant across the frames. Hopefully this common core is unambiguous about the classification of each field as spacetime and/or matter.

Do scalar–tensor theories have a structural common core (Le Bihan and Read 2018; Wolf, Sanchioni and Read 2024, §8.1) that does obey the STM dichotomy? In light of the conformal transformation between the Jordan frame and the Einstein frames, a first guess might be the conformal structure. As far as light rays are concerned, this would be totally fine. The conformal structure consists of the lightcones; these are the same in each frame. Light does not care in which frame it is being represented. However, (a) conformal structure only concerns part of the metric structure, not the full structure of the theory, and, relatedly, (b) conformal structure is not sufficient to unambiguously determine the timelike trajectories of massive particles/observers (cf. Section 3.3.1). Moreover, any such candidate common core that claims to reduce frames (JF), (EF) and (EF') faces the crucial problem that it would be indeterminate whether the action-reaction criterion (Crit. 1) is satisfied, since the coupling of the scalar field to the matter fields $\psi_i$ changes across frames. Retreating by weakening the criterion for matter to merely being a dynamical field would break the STM dichotomy already within GR.

Arguably and surprisingly, it seems that something close to a structural common core—(Wolf, Sanchioni and Read 2024, §8.2) call this an overarching theory, i.e., a synthesis of the entirety of the theoretic structures contained within the frames (JF), (EF), and (EF')—can be found for the particular case of a Brans–Dicke theory, $F = \phi$ and $G = \omega/\phi$, by using integrable Weyl geometry (Lobo 2018; Wolf, Sanchioni and Read 2024). However, this is not possible for the general family of scalar–tensor theories.[24] Furthermore, even in this case, the distinction

24 It is easy to see this by noting that the frame transformation for general $F$ and $G$ cannot be written as a Weyl geometry symmetry transformation $\bar{g} = e^f g$, $\bar{\phi} = \phi + f$ (Lobo 2018).

would be between Weyl-geometrical terms and non-Weyl-geometrical terms, but this is not equivalent to the STM dichotomy.[25]

Without a reduced theory, several interpretational options remain. The analogue of the (strong) interpretationalist may well lament that no common core has been found, but will consider this to be ultimately irrelevant. They may adopt a form of quietism.

**Interpretation 3 (Quietism)**

- **a** **Simple quietism:** There is simply no need to discuss anything beyond invariant structure such as the null geodesics traversed by light—we bury our head in the sand as to what the theory says about trajectories of massive particles.[26]
- **b** **A somewhat more sophisticated quietism:** By analogy with *external sophistication* (Dewar 2019), we make the active, explicit *stipulation* that the different frames/representations represent the same state of affairs.

What does this imply for the STM dichotomy, in light of the previous sections? Simple quietists stamp their foot on the ground and refuse to answer this question. The move that sophisticated quietists make seems coherent only if one gives up a black-and-white STM dichotomy altogether (as being inapplicable in this context), or at least adopts *conventionalism* (Martens and Lehmkuhl 2020b) (i.e., acknowledging that there is a metaphysical distinction between test particles following timelike geodesics or not, but interpreting the theory as not providing an objective matter of fact as to that binary distinction—it is a subjective convention as to which frame one uses to represent the underlying physical state of affairs). Both of these options belong in what Martens and Lehmkuhl (2020b) call the cluster of *strong breakdown* interpretations (of the STM dichotomy in the context of hybrid theories such as scalar–tensor theory). More on this below.

In An Attempt to avoid such a strong breakdown, a remaining alternative one might consider is the following.

**Interpretation 4 (Privilege, or Discrimination[27])** One frame is privileged over the other, in the sense that it represents the one true, metaphysically perspicuous, underlying reality.

- **a** **Fundamentalism:** The other frame is not fundamental, but still valid/real; it emerges from the privileged frame (Martens and Lehmkuhl 2020b).
- **b** **Eliminativism:** The other frame is pathological and does not represent reality in any useful, explanatory, robust manner. We remove it from both our fundamental and non-fundamental ontology.

This is in the spirit of the so-called weak version of interpretationalism and strong version of motivationalism, which allow that theoretical, metaphysical or super-empirical considerations may block the only-invariant-structure-is-real inference (Martens and Read 2021).

Is any of these interpretations a valid subterfuge? No. First, as we will see, there is no representation that obviously wins out. Second, even if this were the case, we have seen (see Table 1 and the opening of this section) that the STM dichotomy breaks down in each frame separately (albeit for different reasons in each case).

25 The JF can be fixed by choosing the Riemannian Frame $(g, 0)$ and the EF choosing the Weyl Frame $(\tilde{g}, \theta)$. In this case $\phi$ in the JF would still be neither spacetime—not part of the Weyl geometry—nor matter. In the EF, $\theta$ would be both part of the spacetime-part of the Weyl geometry—and it would also be matter.

26 This resembles the consubstantiality interpretation in Martens and Lehmkuhl (2020b).

27 See (Le Bihan and Read 2018; Wolf, Sanchioni and Read 2024).

Although this latter point suffices for showing that the privilege interpretations are of no avail if one wants to retain a strict STM dichotomy, we briefly discuss—for those interested in the ontology of scalar–tensor theories independently of that dichotomy—why no representation obviously wins out. On the one hand, the Jordan frame is the frame that naturally falls out of underlying theories such as string theory (i.e., string theory tends to generate non-minimally coupled theories in the low-energy limit) (see Table 2). On the other hand, the Einstein frames have the advantage that their metric and scalar field are not mixed. Score: 1-1.

Could this tie be broken by the "geodesic criterion" (Martens and Lehmkuhl 2020a): for any choice of initial conditions, if force-free test particles (and massive bodies that can be idealized as test particles; and light rays) were around, then, if they would all follow the geodesics of the same affine connection or metric, that connection or metric is the physical spacetime? As shown in Section 3.3.1, in the JF test matter built out of a perfect fluid follows the geodesics of its metric $g_{ab}$, whereas in the EF test matter built out of a perfect fluid does not follow the geodesics of $\tilde{g}_{ab}$ (but rather of $g_{ab}$). This seems to privilege the Jordan frame, as test matter surveys its metric $g_{ab}$ and thereby seems to pick it out as the physical spacetime. However, (1) the relevant representation to contrast the JF with is the EF'; and (2) it is important to keep in mind that the geodesic criterion concerns *force-free* matter—if the test matter does not follow the geodesics of a metric but the reason for that is that it feels a force, this is consistent with that metric being the physical spacetime. In the EF', test matter built out of $\theta$ (which in this representation is matter after all) is indeed force-free and it follows the geodesics of $\tilde{g}_{ab}$ rather than of $g_{ab}$, suggesting that the former is the physical spacetime. On the other hand, while (non-conformally invariant) test matter built out of the regular matter fields $\psi_i$ does not follow the geodesics, the excuse is that they are not force-free (but feel a $\theta$-force). Conclusion: test matter does not privilege one metric over the other.

Finally, could energy conditions break the tie? Also not. In all representations, the strong energy condition is broken—dark energy is after all the umbrella term for anything that accounts for the accelerated expansion of our universe. What about the other three classical energy conditions (weak, null, and dominant)? The Einstein representations satisfy them. Not all solutions of the Jordan representation—with its non-minimal coupling—do satisfy them. A defender of the Jordan frame may flip this around, and retort that it is exactly for that reason that such solutions can be excluded as being unphysical. Although we are skeptical of the validity of this move even in the classical context (see, e.g., Bots (2025)), we know that the classical energy conditions are violated by quantum physics anyway, so energy conditions are not the best tie breaker in this context.

All in all (Table 2), there is no one frame that is obviously privileged.[28] And, at the risk of flogging a dead horse, this would not have saved the STM dichotomy anyway.

| Jordan Frame | Einstein Frames |
|---|---|
| + Naturally produced by string theory | |
| + Geodesic criterion satisfied | + Geodesic criterion satisfied |
| ? Can violate all the classical energy conditions | + Satisfies most classical energy conditions |
| – Metric and scalar field mixed | + Metric and scalar field not mixed |

**Table 2** Reasons against and in favor of privileging each frame.

28 If one would find the Jordan frame's connection sufficiently convincing to privilege that frame, remember that the curvature term in the dynamical equations of that frame violates the philosopher's strong equivalence principle (footnotes 11, 17, and 19).

If one keeps insisting on a black-and-white dichotomy between matter and spacetime, this creates a problem. Hence, we take the first main lesson of this case study to be that there is a strong breakdown (Martens and Lehmkuhl 2020b) of the rigid conceptual dichotomy between spacetime and matter.

**Interpretation 5 (Strong Breakdown, or Gray Spacetime Matter)** Abandon a black-and-white STM dichotomy, in the sense of those categories being mutually exclusive. If one wants to hold on to the spacetime and matter concepts at all, then it is to be understood that a world (described by a scalar–tensor theory) cannot be carved up into physical ingredients that are all either purely an aspect of spacetime structure or purely a form of matter

- **a** **Conventionalism:** There is no objective matter of fact which ontology is correct, i.e., literally interpreting the Jordan frame versus literally interpreting the Einstein frame, each with their different spacetime and matter categorizations. Classifying the scalar field as matter or spacetime is a mere matter of convention.
- **b** **Non-Applicability:** Spacetime and matter are simply not the appropriate categories here. Let us not hold on to them needlessly.

Our analysis of the case study, and the lack of a known common core of the various frames, suggests at least the conventionalist version of this strong breakdown interpretation. We are tempted to go further and adopt the non-applicability interpretation. This conclusion gets reinforced if we ask whether there is another category that does apply, even if spacetime and matter do not do so. We have seen that there is: gravity! In (generally) relativistic contexts, gravity and spacetime are often seen as going hand in hand. As such, the STM dichotomy was implicitly the spacetime/gravity–matter dichotomy. However, although spacetime is not a frame-invariant category in the context of scalar–tensor theories, gravity is (Table 1). The second main lesson of this case study—more positive than the first—is then that gravity, not spacetime, is the appropriate category that remains standing. A consequence of this would be that all fields are grouped into gravitational and non-gravitational—or "other"—fields.

**Interpretation 6 (Black-and-White Gravity, or Conceptual Common Core)** Although spacetime and matter are not mutually exclusive categories in the context of scalar–tensor theories, the conceptual category of gravity does remain applicable, in an objective, frame-independent manner (i.e., Conclusion 5). The gray spacetime/matter dichotomy may be replaced with a black-and-white gravity/non-gravity dichotomy.

A final retort might be that our definition of gravitational fields (Definition 5) refers to matter fields, so in an indirect way the matter category remains after all. Although this may have been problematic for Interpretation 6 in a different context, this does not hinder this interpretation in the context of scalar–tensor theories in any strong sense. The reasons are twofold. First, the dependence on the notion of matter is rather weak, in the sense that categorizing fields as gravitational or non-gravitational does not change depending on whether the scalar field is or is not treated as matter (Section 3.3.2). Second, the dependence on a notion of matter is not a dependence on a black-and-white STM dichotomy, i.e., a notion of matter that excludes spacetime.

# 5 CONCLUSIONS

The status of the STM distinction in GR is controversial. We have argued that it is much more clear-cut that this dichotomy—every field being either an aspect of spacetime structure, or a form of matter, never both, never neither—breaks down in the context of scalar–tensor theories. The main reasons are twofold. First, the categorization of the scalar field, as either matter or spacetime, is not invariant across the different representations/frames of the theory. Second, even within each frame, the dichotomy runs into trouble, although for different reasons in each case: in the Jordan frame, the scalar field is neither spacetime nor matter; in the Einstein frames, the scalar field is either matter or spacetime, depending on which of the two versions of the Einstein representation one considers. We have introduced and analyzed various interpretational options—most importantly: finding a structural common core, quietism, privileging one representation over the other, and a conventionalist form of a breakdown of the dichotomy—and ended up with a stronger breakdown of the dichotomy, i.e., spacetime and matter are simply not the appropriate categories in this context. Something can still be salvaged though. Whereas the concepts of space(time) and gravity are typically understood as having merged, in some sense (Lehmkuhl 2008), with the advent of GR, they can be decoupled again in the context of scalar–tensor theories. The concept of a gravitational field is well-defined in each of the representations of a scalar–tensor theory, and across such representations. If one wants to hold on to one of the old categories, gravity is the one!

# APPENDICES

## A PERFECT FLUID

Take a perfect fluid with stress-energy tensor

$$T^{ab} = \rho u^a u^b + P h^{ab} \tag{1}$$

where $\rho$, $P$ and $u^a$ are the energy density, pressure and four-velocity vector (normalized as $u^a u_a = -1$) respectively and $h^{ab} = g^{ab} + u^a u^b$. The dynamical equations of the fluid in the presence of some force field $f^a$ acting on it can be written as

$$\nabla_a T^{ab} = f^b. \tag{2}$$

This is the case, for instance, for the charged fluid (MAXW-PF). It is easy to show that (see, e.g., Malament (2012)) using (1) and (2)

$$\nabla_a(\rho u^a) + P\nabla_a u^a = -f^b u_b \tag{3}$$

$$(\rho + P)u^a \nabla_a u^b + h^{ab}\nabla_a P = h^b_c f^c \tag{4}$$

which are nothing but the dynamical equations of (PF) in the presence of an external force $f^b$. In the case of $P = 0$, (4) results in a deviation from a geodesic by the given force

$$\rho u^a \nabla_a u^b = h^b_c f^c \tag{5}$$

which for the case of an electromagnetic field is $f^b = -\nabla_a T^{ab}_{\text{EM}} = j^a F_a{}^b$ such that together with a charged fluid (MAXW-PF) $j^a = e u^a$ yields

$$u^a \nabla_a u^b = \frac{e}{\rho} u^a F_a{}^b \tag{6}$$

which is nothing but the Lorentz force. Now, we can apply a conformal transformation to the perfect fluid

$$g_{ab} \to \tilde{g}_{ab} = \phi^2 g_{ab}. \tag{7}$$

In this case, (2) transforms as

$$\tilde{\nabla}_a T^{ab} = f^b + 6\phi^{-1} T^{ab}\tilde{\nabla}_a \phi - \phi^{-1}\tilde{g}^{ab}\tilde{g}_{cd} T^{cd}\tilde{\nabla}_a \phi. \tag{8}$$

We also transform the fluid as follows:

$$u^a \to \tilde{u}^a = \phi^{-1}u^a \qquad \rho \to \tilde{\rho} = \phi^{-4}\rho \qquad P \to \tilde{P} = \phi^{-4}P \tag{9}$$

and further it is useful to redefine the stress-energy tensor as

$$\tilde{T}^{ab} := \phi^{-6}T^{ab} = \tilde{\rho}\tilde{u}^a\tilde{u}^b + \tilde{P}\tilde{h}^{ab} \tag{10}$$

where $\tilde{h}^{ab} = \tilde{g}^{ab} + \tilde{u}^a\tilde{u}^b$. After some calculation (8) transforms to

$$\tilde{\nabla}_a\tilde{T}^{ab} = \phi^{-6}f^b - \tilde{g}^{ab}\tilde{T}\tilde{\nabla}_a \log\phi. \tag{11}$$

where $\tilde{T} := \tilde{g}_{cd}\tilde{T}^{cd}$. Finally, introducing (9) into (11) and following the same procedure as before we obtain

$$\begin{aligned}
&\tilde{\nabla}_a(\tilde{\rho}\tilde{u}^a) + \tilde{P}\,\tilde{\nabla}_a\tilde{u}^a = -(\tilde{\rho} - 3\tilde{P})\tilde{u}^a\tilde{\nabla}_a \log\phi - \phi^{-6}f^b\tilde{u}_b,\\
&(\tilde{\rho} + \tilde{P})\tilde{u}^b\tilde{\nabla}_b\tilde{u}^a + \tilde{h}^{ab}\tilde{\nabla}_b\tilde{P} = (\tilde{\rho} - 3\tilde{P})\tilde{h}^{ab}\tilde{\nabla}_b \log\phi + \phi^{-6}\tilde{h}^a{}_c f^c
\end{aligned} \tag{12}$$

which is nothing but (PF') with an extra force field $f^b$. One could in principle interpret the contribution coming from the conformal transformation as a force field $k^b := (\tilde{\rho} - 3\tilde{P})\tilde{\nabla}^b \log\phi$. However, this is not possible in general (cf. Section 3.3.1), since $\phi$ is not a matter field and therefore there is no transfer of energy-momentum, as is the case for the electromagnetic field generating the force field (6).

## B GEROCH'S FORMULATION OF DYNAMICAL EQUATIONS

A vector bundle consists of a base manifold $M$, a total space $b$, and a projection map $\pi : b \to M$ such that for each point $x \in M$, the pre-image $\pi^{-1}(x)$ is a vector space. These vector spaces are called the fibers of the bundle. Fields can be treated as sections of vector bundles $\Phi : U \subseteq M \to b$ such that $\pi \circ \Phi = \mathrm{Id}$. This means that a field assigns a vector (or more generally a tensor) from the fiber to each point on the base manifold $M$. We interpret the fiber of $x$ as the space of possible field values at this point.

Following Geroch (1996), we may describe a system of partial differential equations on sections. Let $k_A{}^m{}_\alpha$ and $j_A$ be smooth fields on the bundle manifold $b$. Here, the lower-case Latin indices denote tensors on $M$, the Greek indices denote tensors on $b$, and the index $A$ refers to a vector space attached to $b$, whose dimension corresponds to the dimension of the system of equations. For a section $\Phi : U \subseteq M \to b$, consider the equation

$$k_A{}^m{}_\alpha(\nabla\Phi)_m{}^\alpha + j_A = 0, \tag{13}$$

which holds at every $x \in U$. Here $k_A{}^m{}_\alpha$ and $j_A$ are to be evaluated at $\Phi(x)$. The equation is first-order, i.e., linear in the first derivatives of $\Phi$. The number of unknown variables at each point corresponds to the dimension of the fiber. Most known physical systems can be formulated in this general structure, although auxiliary fields may be required for higher-order equations.

Let us apply this formalism to scalar–tensor theory in both the Jordan and Einstein frames. Recall the Jordan system of dynamical equations.

$$\begin{aligned}
&F(\phi)G^{ab} - G(\phi)(\nabla^a\phi\nabla^b\phi - \frac{1}{2}g^{ab}\nabla^c\phi\nabla_c\phi) - \nabla^a\nabla^b F(\phi) + g^{ab}\Box F(\phi) + \frac{1}{2}V(\phi)g^{ab}\\
&\quad = T_M^{ab}[\psi_i, g], 2G(\phi)\Box\phi + G'(\phi)\nabla^c\phi\nabla_c\phi - V'(\phi) + F'(\phi)R = 0,
\end{aligned} \tag{JF}$$

To analyze this as a first-order system, we must introduce the auxiliary equation

$$\phi_a := \nabla_a \phi. \tag{14}$$

Following Geroch (1996), we also think of the curvature tensor as the derivative of the derivative operator. Written in the form of Eq. 13, the Einstein Equation in the Jordan frame has a dynamical sector (the $\nabla\Phi$ part) containing fields $\nabla_a$ and $\phi_b$. The other terms of this equation are source terms. The Klein–Gordon equation in the Jordan frame also has as dynamical fields $\nabla_a$ and $\phi_b$. Observe the mixing between the scalar field and the (spacetime) "connection field" $\nabla_a$.

Next, recall the Einstein Frame system of equations of the scalar–tensor theory.

$$R^{ab}[\tilde{g}] - \frac{1}{2}R[\tilde{g}]\tilde{g}^{ab} - \tilde{\nabla}^a\theta\tilde{\nabla}^b\theta + \frac{1}{2}\tilde{g}^{ab}\tilde{g}^{cd}\tilde{\nabla}_c\theta\tilde{\nabla}_d\theta + U(\theta)\tilde{g}^{ab} = \frac{1}{F^3}T_M^{ab}[\psi_i, F^{-1}\tilde{g}]. \tag{15}$$

$$\tilde{\Box}\theta - U'(\theta) = \frac{F'}{2F^4}\tilde{g}_{ab}T_M^{ab}[\psi_i, F^{-1}\tilde{g}] \qquad \text{(EF)} \tag{16}$$

To analyze this as a first-order system (considering that the Klein-Gordon equation still contains a second-order derivative of the scalar field), we must introduce the auxiliary equation

$$\theta_a := \nabla_a \theta. \tag{17}$$

The only dynamical field in the Einstein Equation in the Einstein frame is then $\tilde{\nabla}_a$. The Klein-Gordon equation in the Einstein frame only has the dynamical field $\theta_a$. Therefore, the mixing between the scalar field and spacetime degrees of freedom has disappeared.

# C FRAME TRANSFORMATIONS

In this appendix, we derive the Einstein-frame equations of motion of a general scalar–tensor theory.

## The Einstein equation

Start with the generalized Einstein equation:

$$F(\phi)(R_{ab} - \frac{1}{2}Rg_{ab}) - \nabla_a\nabla_b F + \Box F g_{ab} - G(\phi)(\nabla_a\phi\nabla_b\phi - \frac{1}{2}g_{ab}\nabla_c\phi\nabla^c\phi) + \frac{1}{2}Vg_{ab} = T_{ab}. \tag{18}$$

For any smooth, strictly positive function $\Omega$ we can conformally transform the metric as

$$g_{ab} \to \tilde{g}_{ab} := \Omega^2 g_{ab}.$$

The Ricci tensor and Ricci scalar transform as follows (see Appendix D in Wald (1984))

$$\begin{aligned}\tilde{R}_{ab} = R_{ab} - 2\nabla_a\nabla_b\log\Omega - g_{ab}g^{cd}\nabla_c\nabla_d\log\Omega + 2(\nabla_a\log\Omega)\nabla_b\log\Omega \\ - 2g_{ab}g^{cd}(\nabla_c\log\Omega)\nabla_d\log\Omega\end{aligned} \tag{19}$$

$$\tilde{R} = \Omega^{-2}(R - 6g^{ab}\nabla_a\nabla_b\log\Omega - 6g^{ab}(\nabla_a\log\Omega)\nabla_b\log\Omega). \tag{20}$$

where $\tilde{R}_{ab}$ and $\tilde{R}$ are the Ricci tensor and Ricci scalar with respect to the new metric $\tilde{g}_{ab}$. In order to eliminate the second-derivative terms $-\nabla_a\nabla_b F + \Box F g_{ab}$, we can make a judicious choice by setting $\Omega := F^{1/2}$ such that

$$\nabla_b\log\Omega = \nabla_b\log(F^{1/2}) = \frac{\nabla_b F}{2F} \tag{21}$$

$$\nabla_a\nabla_b\log\Omega = \nabla_a\nabla_b\log(F^{1/2}) = \frac{1}{2F}(\nabla_a\nabla_b F - \frac{1}{F}\nabla_a F\nabla_b F). \tag{22}$$

Using (19), (20), (21), and (22), we may express the Einstein tensor as

$$G_{ab} = \tilde{G}_{ab} + \frac{1}{F}\nabla_a\nabla_b F - \frac{3}{2F^2}\nabla_a F\nabla_b F - \frac{1}{F}g_{ab}g^{cd}\nabla_c\nabla_d F + \frac{3}{4F^2}g_{ab}g^{cd}\nabla_c F\nabla_d F. \quad (23)$$

Then, (18) becomes

$$F\tilde{G}_{ab} - \frac{3}{2F}\nabla_a F\nabla_b F + \frac{3}{4F}g_{ab}g^{cd}\nabla_c F\nabla_d F - G(\nabla_a\phi\nabla_b\phi - \frac{1}{2}g_{ab}\nabla_c\phi\nabla^c\phi) + \frac{1}{2}g_{ab}V = T_{ab}. \quad (24)$$

Notice that the terms involving double derivatives of $F$ cancel out here. Next, recall that $F = F(\phi)$, so that

$$\nabla_a F = F'\nabla_a\phi, \quad (25)$$

where the prime denotes differentiation with respect to $\phi$. We obtain

$$F\tilde{G}_{ab} - (\frac{3F'^2}{2F} + G)(\tilde{\nabla}_a\phi\tilde{\nabla}_b\phi - \frac{1}{2}\tilde{g}_{ab}\tilde{g}^{cd}\tilde{\nabla}_c\phi\tilde{\nabla}_d\phi) + \frac{1}{2F}V\tilde{g}_{ab} = T_{ab}[\psi_i, F^{-1}\tilde{g}_{ab}], \quad (26)$$

using the fact that $\tilde{\nabla}_a\phi = \nabla_a\phi$. We redefine the scalar field as

$$\phi \to \theta, \quad \text{such that} \quad \frac{d\theta}{d\phi} = \sqrt{\frac{G}{F} + \frac{3F'^2}{2F^2}}. \quad (27)$$

Note that

$$\tilde{\nabla}_a\theta = \frac{d\theta}{d\phi}\tilde{\nabla}_a\phi = \sqrt{\frac{G}{F} + \frac{3F'^2}{2F^2}}\tilde{\nabla}_a\phi. \quad (28)$$

Then the Einstein equation in the Einstein frame in terms of $\theta$ is

$$\tilde{G}_{ab} - \tilde{\nabla}_a\theta\tilde{\nabla}_b\theta + \frac{1}{2}\tilde{g}_{ab}\tilde{g}^{cd}\tilde{\nabla}_c\theta\tilde{\nabla}_d\theta + \frac{1}{2F^2}V\tilde{g}_{ab} = \frac{1}{F}T_{ab}. \quad (29)$$

Taking the trace with respect to $\tilde{g}_{ab}$ gives, with $T := \tilde{g}^{ab}T_{ab}$ (note that this differs from the Jordan-frame trace $g^{ab}T_{ab} = FT$),

$$\tilde{R} - \tilde{g}^{ab}\tilde{\nabla}_a\theta\tilde{\nabla}_b\theta - \frac{2V}{F^2} = -\frac{T}{F}. \quad (30)$$

## The Klein–Gordon equation

We now turn to the generalized Klein–Gordon equation:

$$2G\Box\phi + G'\nabla^c\phi\nabla_c\phi - V'(\phi) + F'R = 0. \quad (31)$$

From Appendix D in Wald (1984), we know that a dual vector field $\omega_b$ transforms under $\Omega$ as

$$\tilde{\nabla}_a\omega_b = \nabla_a\omega_b - \left(\delta^c_a\nabla_b\log\Omega + \delta^c_b\nabla_a\log\Omega - g_{ab}g^{cd}\nabla_d\log\Omega\right)\omega_c. \quad (32)$$

Now, taking again $\Omega := F^{1/2}$, we can express the box operator in terms of the transformed quantities as

$$\begin{aligned}
\Box\phi &= g^{ab}\nabla_a(\tilde{\nabla}_b\phi) && (33)\\
&= F\tilde{g}^{ab}(\tilde{\nabla}_a\tilde{\nabla}_b\phi) + g^{ab}\left(\delta^c_a\nabla_b\log\Omega + \delta^c_b\nabla_a\log\Omega - g_{ab}g^{cd}\nabla_d\log\Omega\right)\nabla_c\phi && (34)\\
&= F\tilde{\Box}\phi + \frac{F'}{2F}g^{ab}\nabla_a\phi\nabla_b\phi + \frac{F'}{2F}g^{ab}\nabla_b\phi\nabla_a\phi - \frac{2F'}{F}g^{cd}\nabla_d\phi\nabla_c\phi && (35)\\
&= F\tilde{\Box}\phi - F'\tilde{\nabla}^c\phi\tilde{\nabla}_c\phi && (36)
\end{aligned}$$

Recall the transforming of the Ricci scalar under a conformal transformation, from Eq. 20. Then

$$R = F\tilde{R} + \frac{3}{F}g^{ab}(\nabla_a\nabla_b F - \frac{1}{F}\nabla_a F\nabla_b F) + \frac{3}{2F^2}g^{ab}\nabla_a F\nabla_b F$$
$$= F\tilde{R} + \frac{3}{F}g^{ab}\nabla_a\nabla_b F - \frac{3}{2F^2}g^{ab}\nabla_a F\nabla_b F. \qquad (37)$$

We will use that

$$\nabla_a\nabla_b F = F''\nabla_a\phi\nabla_b\phi + F'\nabla_a\nabla_b\phi. \qquad (38)$$

The transformed Klein–Gordon equation is then

$$2G\left(F\tilde{\Box}\phi - F'\tilde{g}^{ab}\tilde{\nabla}_a\phi\tilde{\nabla}_b\phi\right) + FG'\tilde{g}^{ab}\tilde{\nabla}_a\phi\tilde{\nabla}_b\phi - V' +$$
$$FF'\tilde{R} + \frac{3F'}{F}g^{ab}\left(F''\nabla_a\phi\nabla_b\phi + F'\nabla_a\nabla_b\phi\right) - \frac{3F'}{2F^2}g^{ab}\nabla_a F\nabla_b F = 0$$
$$(2FG + 3F'^2)\tilde{\Box}\phi - (2F'G - FG' - 3F'F'' + \frac{9}{2}\frac{F'^3}{F})\tilde{g}^{ab}\tilde{\nabla}_a\phi\tilde{\nabla}_b\phi - V' + FF'\tilde{R} = 0. \qquad (39)$$

Again, redefine the scalar field as

$$\phi \to \theta, \qquad \text{such that } \frac{d\theta}{d\phi} = \sqrt{\frac{G}{F} + \frac{3F'^2}{2F^2}} =: \sqrt{\Delta}. \qquad (40)$$

We have

$$\tilde{\Box}\theta = \tilde{g}^{ab}\tilde{\nabla}_a(\frac{d\theta}{d\phi}\tilde{\nabla}_b\phi), \qquad (41)$$

which implies that

$$\tilde{\Box}\phi = \frac{1}{\sqrt{\Delta}}\tilde{\Box}\theta - \frac{1}{\sqrt{\Delta}}\frac{d^2\theta}{d\phi^2}\tilde{g}^{ab}\tilde{\nabla}_a\phi\tilde{\nabla}_b\phi. \qquad (42)$$

It will be useful to compute

$$\frac{d^2\theta}{d\phi^2} = \frac{1}{2\sqrt{\Delta}}\frac{1}{F^4}\left(F^3G' - F^2F'G + 3F^2F'F'' - 3FF'^3\right). \qquad (43)$$

The Klein–Gordon equation then becomes

$$2F^2\Delta\left(\frac{1}{\sqrt{\Delta}}\tilde{\Box}\theta - \frac{1}{\sqrt{\Delta}}\frac{d^2\theta}{d\phi^2}\frac{1}{\Delta}\tilde{g}^{ab}\tilde{\nabla}_a\theta\tilde{\nabla}_b\theta\right) -$$
$$\left(2F'G - FG' - 3F'F'' + \frac{9F'^3}{2F}\right)\frac{1}{\Delta}\tilde{g}^{ab}\tilde{\nabla}_a\theta\tilde{\nabla}_b\theta - V' + FF'\tilde{R} = 0$$
$$2F^2\sqrt{\Delta}\tilde{\Box}\theta - FF'\tilde{g}^{ab}\tilde{\nabla}_a\theta\tilde{\nabla}_b\theta - V' + FF'\tilde{R} = 0. \qquad (44)$$

Using the trace of the Einstein equation (Eq. 30), this simplifies to

$$\tilde{\Box}\theta - \frac{1}{\sqrt{\Delta}}\frac{1}{2F^2}V' + \frac{1}{\sqrt{\Delta}}\frac{F'}{F^3}V = \frac{1}{\sqrt{\Delta}}\frac{F'}{2F^2}T. \qquad (45)$$

Finally, we redefine

$$' = \frac{d}{d\phi} \quad \to \quad ' := \frac{d}{d\theta}, \qquad \text{using } \frac{d}{d\phi} = \frac{d\theta}{d\phi}\frac{d}{d\theta} = \sqrt{\Delta}\frac{d}{d\theta}. \qquad (46)$$

This gives

$$\tilde{\Box}\theta - \frac{1}{2F^2}V' + \frac{F'}{F^3}V = \frac{F'}{2F^2}T. \qquad (47)$$

Setting $U := V/2F^2$, we get

$$\tilde{\Box}\theta - U'(\theta) = \frac{F'}{2F^4}\tilde{g}_{ab}T_M^{ab}[\psi_i, F^{-1}\tilde{g}] \tag{48}$$

## Stress-energy tensor

We transform the Jordan-frame matter tensor as

$$\tilde{T}_M^{ab} := F^{-3}\, T_M^{ab}[\psi_i,\, F^{-1}\tilde{g}]. \tag{49}$$

Suppose we have matter conservation in the Jordan frame

$$\nabla_a T_M^{ab} = 0. \tag{50}$$

For any symmetric $T^{ab}$ and $\tilde{g}_{ab} = \Omega^2 g_{ab}$, we have (see Eq. D.20 from Wald (1984))

$$\tilde{\nabla}_a(\Omega^s T^{ab}) = \Omega^s \nabla_a T^{ab} + (s+n+2)\,\Omega^{s-1}T^{ab}\nabla_a\Omega - \Omega^{s-1}g^{ba}\,T\,\nabla_a\Omega, \tag{51}$$

with $n$ the spacetime dimension.

We set $n = 4$, $s = -6$ (so that $s + n + 2 = 0$). Using $\nabla_a T_M^{ab} = 0$ we obtain

$$\tilde{\nabla}_a\tilde{T}_M^{ab} = -\,\Omega^{-7}\,g^{ba}\,T_M\,\nabla_a\Omega \tag{52}$$

$$= -\,\frac{1}{2F}\,\tilde{T}_M\,\tilde{\nabla}^{\,b}F, \tag{53}$$

where in the last equality we used $g^{ba} = F\tilde{g}^{ba}$, $T_M = g_{cd}T_M^{cd} = F^2\tilde{T}_M$, and $\tilde{g}^{ba}\nabla_a\Omega = \tilde{\nabla}^{\,b}\Omega = \frac{1}{2F^{1/2}}\tilde{\nabla}^{\,b}F$.

Recall that we defined the scalar stress-energy tensor as

$$T_\theta^{ab} = \tilde{\nabla}^a\theta\,\tilde{\nabla}^b\theta - \frac{1}{2}\tilde{g}^{ab}(\tilde{\nabla}\theta)^2 - U(\theta)\,\tilde{g}^{ab}. \tag{54}$$

We compute:

$$\tilde{\nabla}_a T_\theta^{ab} = (\tilde{\Box}\theta)\,\tilde{\nabla}^b\theta + \tilde{\nabla}^a\theta\,\tilde{\nabla}_a\tilde{\nabla}^b\theta - \frac{1}{2}\tilde{\nabla}^{\,b}\big((\tilde{\nabla}\theta)^2\big) - U'(\theta)\,\tilde{\nabla}^{\,b}\theta$$

$$= \big(\tilde{\Box}\theta - U'(\theta)\big)\,\tilde{\nabla}^{\,b}\theta. \tag{55}$$

Using the (EF') scalar equation

$$\tilde{\Box}\theta = \frac{F'}{2F}\,\tilde{g}_{cd}\,\tilde{T}_M^{cd}[\tilde{\psi}_i, \tilde{g}, \theta] + U'(\theta), \tag{56}$$

we get

$$\tilde{\nabla}_a T_\theta^{ab} = \frac{F'}{2F}\,\tilde{T}_M\,\tilde{\nabla}^{\,b}\theta = \frac{1}{2F}\,\tilde{T}_M\,\tilde{\nabla}^{\,b}F. \tag{57}$$

Adding the matter and scalar contributions yields the desired Einstein–frame conservation law

$$\tilde{\nabla}_a\big(\tilde{T}_M^{ab}[\psi_i, \tilde{g}, \theta] + T_\theta^{ab}[\theta, \tilde{g}]\big) = 0. \tag{58}$$

# ACKNOWLEDGMENTS

We would like to thank the Utrecht Philosophy of Astronomy & Cosmology (UPAC) research group, and especially Sanne Vergouwen, for valuable feedback on an earlier draft of this paper. This paper has also benefited from discussions with the audiences at the 2nd History and Philosophy of Cosmology Conference in Milan (2024), the Lichtenberg Research Seminar in History and Philosophy of Physics in Bonn (2025), the Spacetime Matters Conference in Utrecht (2025), the 22nd European Conference on Foundations of Physics in Gdansk (2025), the BSPS annual conference in Glasgow (2025), and the 10th biennial meeting of the European Philosophy of Science Association in Groningen (2025).

# FUNDING INFORMATION

This project has received funding from the European Union's Horizon Europe ERC programme (ERC Starting Grant Agreement No 101076402 – COSMO-MASTER). Views and opinions expressed are, however, those of the author(s) only and do not necessarily reflect those of the European Union or the European Research Council Executive Agency (granting authority). Neither the European Union nor the granting authority can be held responsible for them.

# COMPETING INTERESTS

The authors have no competing interests to declare.

# AUTHOR AFFILIATIONS

**Antonio Ferreiro** orcid.org/0000-0001-7448-9046
Freudenthal Institute, Utrecht University, NL; Descartes Centre for the History and Philosophy of the Sciences and the Humanities, Utrecht University, NL

**Alex Fleuren** orcid.org/0009-0002-2501-8920
Descartes Centre for the History and Philosophy of the Sciences and the Humanities, Utrecht University, NL

**Niels C. M. Martens** orcid.org/0000-0002-2839-1387
Freudenthal Institute, Utrecht University, NL; Descartes Centre for the History and Philosophy of the Sciences and the Humanities, Utrecht University, NL; Black Hole Initiative, Harvard, US